\documentclass{sig-alternate}

\usepackage{endnotes}
\usepackage{times}
\usepackage{balance}
\usepackage{flushend}
\usepackage{amsfonts}
\usepackage{amssymb}
\usepackage{amscd}
\usepackage{amsmath}
\usepackage{array}
\usepackage{epstopdf}
\usepackage{algorithm, algorithmic}
\usepackage{multirow}
\usepackage{graphicx}
\usepackage{subfigure}
\usepackage{url}
\makeatletter
\g@addto@macro{\UrlBreaks}{\UrlOrds}
\makeatother
\usepackage[hyphenbreaks]{breakurl}
\usepackage{tabularx}

\newcommand{\ignore}[1]{}
\newcommand{\revised}[1]{}
\newcommand\comment[1]{}

\usepackage{color}
\usepackage{listings}
\usepackage{soul}
\usepackage{graphicx}
\begin{document}

\title{Acoustic Fingerprinting Revisited: Generate Stable Device ID Stealthy with Inaudible Sound}

\numberofauthors{1}
\author{
\alignauthor Zhe Zhou, Wenrui Diao, Xiangyu Liu, Kehuan Zhang\\
       \affaddr{Department of Information Engineering} \\
       \affaddr{The Chinese University of Hong Kong}
}

%\numberofauthors{1}
%\author{
%\alignauthor Zhe Zhou, Wenrui Diao, Xiangyu Liu, Kehuan Zhang
%       \affaddr{Department of Information Engineering} \\
%       \affaddr{The Chinese University of Hong Kong}
%}

\maketitle
\begin{abstract}

The popularity of mobile device has made people's lives more convenient, but threatened people's privacy at the same time. As end users are becoming more and more concerned on the protection of their private information, it is even harder to track a specific user using conventional technologies. For example, cookies might be cleared by users regularly. Apple has stopped apps accessing UDIDs, and Android phones use some special permission to protect IMEI code. To address this challenge, some recent studies have worked on tracing smart phones using the hardware features resulted from the imperfect manufacturing process. These works have demonstrated that different devices can be differentiated to each other. However, it still has a long way to go in order to replace cookie and be deployed in real world scenarios, especially in terms of properties like uniqueness, robustness, etc. In this paper, we presented a novel method to generate stable and unique device ID stealthy for smartphones by exploiting the frequency response of the speaker. With carefully selected audio frequencies and special sound wave patterns, we can reduce the impacts of non-linear effects and noises, and keep our feature extraction process un-noticeable to users.  The extracted feature is not only very stable for a given smart phone speaker, but also unique to that phone. The feature contains rich information that is equivalent to around 40 bits of entropy, which is enough to identify billions of different smart phones of the same model.  We have built a prototype to evaluate our method, and the results show that the generated device ID can be used as a replacement of cookie.

%Users have already realized the consequence of privacy leakage resulted from being tracing by the service provider, which drive them to block or clear cookie. Substitution such as IMEI can neither escape from being limited or cut off. Fingerprinting users becomes gradually difficult for service provider, which however never stops their seeking for new techniques. We found that the frequency response feature of speaker in the smart phone differs from each other because of their physical variations. We designed a scheme that measures the higher frequency range response feature of the phone to generate device ID as a kind of substitution to cookie. Measuring the response feature requires acoustic stimulation, which may attract user's attention usually. We chose the higher frequency stimulation, which can hardly be perceived by human being. Besides the unawareness, the entropy is more abundant resulted from looser quality control in this frequency range. The scheme is evaluated, where the result is positive at almost all metrics including stability, anti-interference, and entropy.

\end{abstract}

\section{introduction}
\label{sec:intro}

Smart phone is playing an increasingly important role in our daily lives, including both work and personal entertainment, which makes the security of smart phones a very important and urgent problem, especially the protection of user privacy. Smart phone sales are experiencing nearly 40\% year on year increasing reported by IDC\cite{IDC}. However, according to F-secure, a continued 49\% raising of mobile threat was witnessed in the last quarter, and 91.3\% of them targeted at Android platform, the most popular mobile operating system today\cite{F-secure}. Different from traditional desktop PCs, smart phones often contain more private and sensitive information, like SMS, contacts, location, etc. And studies showed that such sensitive data is the major reason why smart phones are so attractive to attackers~\cite{wei2012android}.

Fortunately, people are becoming better educated to know how to protect their privacy.  Statistics from Pew Internet Project shows that almost 90\% of adult Internet users have taken steps to avoid surveillance by other people or organizations, like clearing cookies, encrypting email, and using an alias~\cite{clearcookie}. To attract users, major browsers now support various privacy protection features, like ``Don't Track'', third party cookie disabling, etc. Governments and organizations are also working on laws to protect users' privacy.

%Cookie is a piece of data stored in client set by the server, thus, the service provider, to make the stateless HTTP sessional and thereby track user's browsering history. The web becomes colourful because of the inventing of cookie. Because of cookie, you are not asked to enter the account and password each time you click a hyperlink. It is also cookie that shopping cart in the ebay becomes available. Besides the convenience it takes to us, cookie also invoked some discussion about security and privacy. For example, a 3rd-party advertising companies may place images or web bugs into every advertisement bearing pages. Each time you visit the page, the image will also be downloaded from the ad web site, during which the cookie will also be set. The targeted user will all be marked with this specific cookie without user's permission\cite{jackson2006protecting}.

However, being able to track users is really useful and important in many legitimate applications. So, it is not surprising to see that many big companies declare plans to give up using cookie on one side, but also work on new tracking technologies on the other side~\cite{googlecookie}. There are also many studies on the stop-tracking and new tracking technologies in the academia world~\cite{roesner2012detecung,mayer2012third,soltani2010flash,krishnamurthy2009privacy,toubiana2010adnostic,krishnamurthy2006generating,nikiforakis2013cookieless}.

%Don't Track me feature, 3rd-party cookie disabling, even cookie disabling has been adopted to anti privacy leakage resulted form cookie. The service provider has already realized the drawback of cookie and has been seeking substitution to cope with users who limited cookie's usage. They have already found that device ID is a substitution to cookie\cite{UUID}. Specifically, device fingerprint is a more straightforward means of tracking comparing to cookie which is set intentionally and can be easily removed. However defender is also give response to this kind of method, thus, accessing to some fingerprint such as unique ID is becoming more and more difficult. For example, in Android system, accessing to IMEI will be blocked by default if a MDM has been installed on the device\cite{IMEI_block}.

Among these new tracking technologies, some suggested to use device ID to substitute cookies~\cite{UUID}, mainly because that device ID is more straightforward and cannot be wiped or reset easily. Typically, many things can be used as device ID, such as UDID (Unique Device ID) from Apple, IMEI for general mobile phones, Android ID for Android phones, MAC addresses of Wi-Fi and Ethernet network interfaces or Bluetooth modules~\cite{takeda2012user}, and so on. Some recent researches also suggested to construct device ID using hardware features resulted from imperfect manufacture process, like the accelerometers~\cite{dey2014accelprint} and speakers~\cite{das2014fingerprinting}.

But each of these solutions has its own limitations which make it hard to replace the traditional user tracking approach based on cookies. On one hand, system vendors can easily block the access of a device ID by removing relevant APIs, and on the other hand, some newly discovered device ID is not mature enough to be deployed in real world production scenarios. For example, Apple ceased the use of UDID recently~\cite{appleUDID}, and on Android, accessing to IMEI requires a special permission that could be revoked by Google if necessary (actually, Google made changes to Android permission system from time to time, and recently, they just took back the permission on SD card writing~\cite{android44sdwrite}, so there is no guarantee that they would not take back permission related to IMEI and other possible device IDs). For newly discovered device IDs, like the one extracted from accelerometers and speakers, the false positive rate is still too high and they are not stable and robust enough to give each device an unique ID (more details are given in section~\ref{sec:rel}).

So, in this paper we propose another device ID generation method that could reach the requirement that a cookie replacement should do: uniqueness, robustness, and stealthy. Our basic idea is hardware-based identification on smart phone by leveraging frequency response of speaker, while our technical is, however, totally different from previous work, which improved the final results dramatically.

One of our fundamental differences to previous work is the use of high frequency sound. In previous work~\cite{das2014fingerprinting}, a piece of music is played, which falls to the frequency range normally lower than 10~kHz, thus can be easily heard by the smart phone owner. What's more, majority of our environmental noises also fall into this range, which makes the feature extraction difficult and unstable.

By contrast, our method uses audio frequency higher than 14~kHz, which is chosen after careful studies of various factors, including the environment noise, characteristics of human hearing, as well as the manufacturing technology of speakers. For example, as shown by our experiments in section~\ref{sec:eval}, in most cases, there are less noise in higher frequency range. What's more, studies of human hearing indicate that our ears are much less sensitive at sound with higher frequency, which means that people can easily hear a sound with 4~kHz at 30~\texttt{dB}, but is hard to perceive another 16~kHz sound at the same 30~\texttt{dB} (more details are given in section~\ref{sec:overview}).

More importantly, we found that speakers perform more diversely at higher frequency range, which helped us be able to get unique feature for each of them with negligible false positive and false negative rates. Ideally, we would expect each speaker perform in the same way: output every frequency equally without any attenuation. However, this is impossible in real world, so speaker manufacturers have to make trade-offs among the cost, manufacturing technology, and the perception of human ears. As mentioned above, people are more sensitive to low frequency audio, so the speaker manufacturers focused on the optimization at lower frequency range first, and optimize higher frequency range later only if cost/budget permits. As the result, it is not surprising that the frequency response curves of the same products are similar at lower frequency range, but differ to each other dramatically at higher frequency range (more details will be given in section~\ref{sec:overview}).

Another fundamental difference to previous work is that we construct audio stimulus pattern carefully to minimize the impacts of non-linear characteristics of speaker and background noises. Instead of playing a piece of random chosen music, as was done in previous work, we choose to output a stable combination of about seventy different frequencies, and later when extracting features, only analyze response at these frequency points. So, noises not on those frequency points can be filtered, but more importantly, the speaker can work in a stable state in which its features can be exposed steadily and completely. We believe that such design is crucial to get unique and robust device ID.

\textbf{Contributions.} We summarize our contributions as follows:
\begin{itemize}
  \item{We carefully analyzed many different factors that could affect the construction of unique and robust device ID from mobile phone speakers, and proposed to use high frequency sound with special frequency pattern as stimulation to speakers, which not only can make the whole process unnoticeable by the smart phone owners, but also can minimize the impact of background noises and non-linear features.}
  \item{We developed novel algorithms to extract and match features from the recorded speaker response, which is built on self-correlation and cross-correlation functions, instead of using complex machine learning algorithm. We also developed method to estimate the potential false positive and false negative rate.}
  \item{We built a prototype and performed a comprehensive evaluation over the proposed method, and the results show that the extracted device ID is very stable, with negligible false positive and false negative rates. }
%\item{We evaluate the scheme and conclude that sufficient entropy can be accumulated by exploiting this feature.}
%\item{We argue that this feature can be matched with the existed fast matching algorithm such as LSH.}
%\item{We propose that the unconstrained speaker permission may leak user's location information with user's identity.}
\end{itemize}

\textbf{Roadmap.} The rest of the paper is organized as follows. We list required assumptions and adversary models in Section~\ref{sec:adversary} and then give an overview of our proposed method in section~\ref{sec:overview}. The details of our design in given in section~\ref{sec:impl}, followed by a comprehensive evaluation of the proposed design in terms of different metrics in Section~\ref{sec:eval}. Section~\ref{sec:application} presents some real world application cases in. We compare our work with prior ones in Section~\ref{sec:rel}, and discussed the potential limitations in Section~\ref{sec:discuss}. Section~\ref{sec:conclusion} concludes the paper.

\section{Adversary Model}
\label{sec:adversary}

This section describe the assumptions required to extract device ID from smart phone speakers, and the potential adversary/application scenarios our method may be applicable.

\subsection{Application Scenarios}
\label{adversarymodel}

As a device fingerprinting technology, the method to be investigated in this paper is pretty neutral, and its only purpose to extract some features from the sound played by smart phone speakers. There are two typical application scenarios: self-fingerprinting and cross-fingerprinting. In self-fingerprinting, an application is trying to get device ID of the smart phone on which it is running, and in cross-fingerprinting, application on one smart phone is trying to get device ID of another smart phone (with the help of an app on that phone which is periodically playing specially crafted audio).

The extracted device ID itself can have many useful applications. For example, it can be used to replace cookie to accurately trace an end user by online advertisers in order to deliver targeted advertisements. It can also be used to in-door tracking and tracking stolen smart phone to support self-destruct. More details will be given in Section~\ref{sec:application}.

\subsection{Assumptions}
\label{assumptions}

The device fingerprinting process actually contains three steps: play the specially crafted audio, record the speaker output, and transmit the preprocessed feature to server. These three steps can be mapped to three different operations or permissions: play audio, access microphone, and access Internet.

\begin{itemize}
  \item{Play audio: According to current Android permission mechanism, playing audio does not require any permission.}
  \item{Access to microphone: This is the only necessary permission required by our proposed method, since we have to record the speaker output. However, depending on the specific application scenario, the microphone permission could locate on the same phone that playing the audio (i.e., self-fingerprinting), or on a different phone (cross-fingerprinting). }
  \item{Access to Internet: This permission is unnecessary and can be bypassed using an existing vulnerability mentioned in~\cite{zhou2013identity} by appending the data to a \texttt{GET} request of stock browsers. The size of each extracted feature never exceeds 1~KB, so the length limitation of GET request is also not a problem.}

\end{itemize}

\section{Overview}
\label{sec:overview}

In this section we introduce the reason why to study sound acoustic fingerprinting of mobile devices though some related work already existed, and briefly describe the technical background of our approach.

\subsection{Three Goals to Be Achieved}

We believe that any device fingerprinting technology, in order to be a substitution of cookie, should achieve following three goals simultaneously: \textit{uniqueness}, \textit{robustness}, and \textit{stealthiness}. In terms of {\bf uniqueness}, the fingerprints generated for different devices should be different enough to each other, otherwise there would be serious usability problem (imagine that two different users share an identical cookie). {\bf Robustness} means the fingerprints generation method should be able to generate a consistent fingerprints for the same device at different time and under different scenarios. The last goal, {\bf stealthiness}, require the fingerprints generation process should be unnoticeable by device owners.

%%% it seems better to put it into evaluation section.
%To better evaluate the performance of our method, we propose to use following quantitative indicators: \textit{false positive} rate, \textit{false negative} rate, and \textit{error} rate. A false positive case happens when a new device A is recognized as an existing device B, and a false negative case arises when an existing device B is regarded as a new device. We say an error is made when an existing device B is recognized as another {\em existing} device.

{\bf Limitations of existing solutions.}  When considering above goals, we found that existing solutions have various limitations. For example, the work done in~\cite{das2014fingerprinting} needs to play some audible music, which make it hard to achieve ``stealth'' goal. In another work that uses accelerometers to track user, there would always be at least 1 device out of 107 wrongly identified, which may not be accurate enough for cookie based applications in real world~\cite{dey2014accelprint}. More details will be given in related work section ~\ref{sec:rel}.

\subsection{Our Key Techniques}
Our key techniques could be described in a single sentence: use microphone to record the output from device speaker stimulated by high frequency audio wave with some special pattern. However, it requires more words to explain the rationale behind and how uniqueness, robustness, and stealthiness are achieved with these techniques.

\subsubsection{Be Stealthy with High Frequency Audio}
\begin{figure}[h]
\centering
\includegraphics[width=0.5\textwidth]{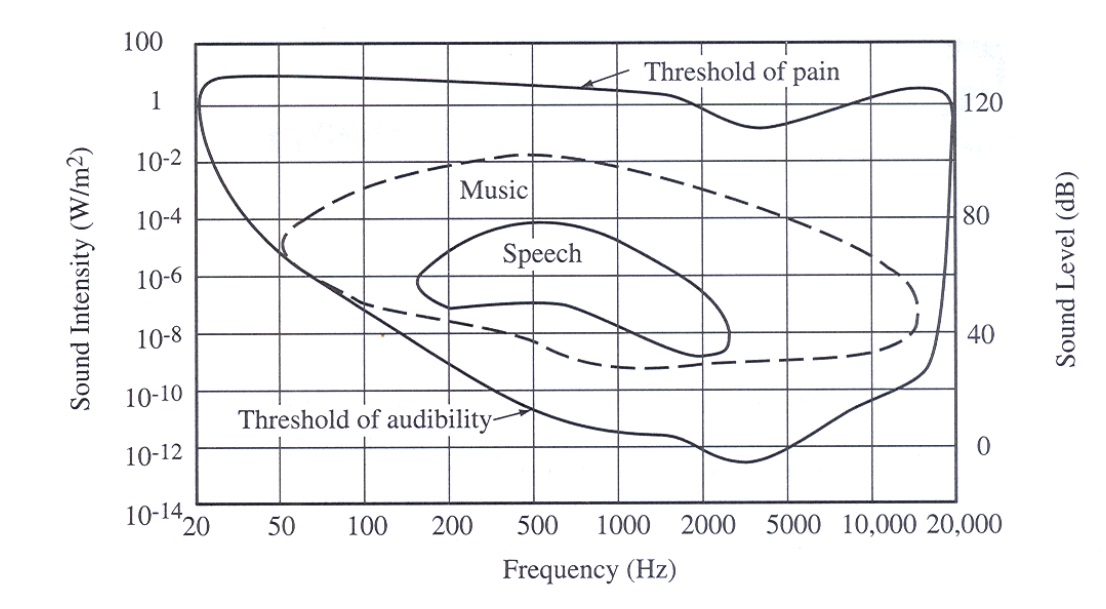}
\caption{Sound Pressure Level of Human over Frequency.}
\label{spl}
\end{figure}

Common sense tells us that human being cannot hear all voice generated by the world. For example, infrasonic wave produced by earthquake doesn't make any feeling to human but can be detected by machines, which play an important role in the disaster forecasting. Ultrasonic, possesses similar attributes. Figure.~\ref{spl} shows how is human's hearable zone~\cite{SPL}.

Most people are sensitive from some hundreds Hz to some thousands Hz and can only feel little lower than 200 or higher than 15 kHz if the sound is as loud as what generated by the phone. In other words, you can hear almost nothing if your cell-phone is playing a clip of music of which spectrum is null between 200 and 15 kHz.

\subsubsection{Be Unique with High Frequency Audio}
\begin{figure}[h]
\centering
\includegraphics[width=0.45\textwidth]{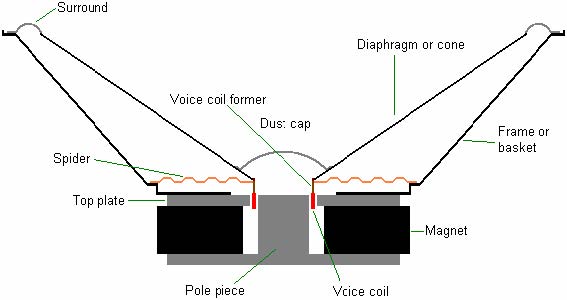}
\caption{Sectional View to Speaker Driver.}
\label{speaker}
\end{figure}

Inside each speaker driver, a flexible cone attached with a coil of wire is mounted on the suspension, which allow it move freely inside the magnet. The coil, passed with electrical currents, creates a varying magnetic field that react with the fixed magnet and drive the cone to fluctuate according to the currents ~\cite{brown2006linear}. Figure.~~\ref{speaker} illustrates the structure of the speaker~\cite{brown2006linear}.

High-end speaker systems may contain more than a single driver to let each driver focused on each frequency band and enhance the quality thereby, because that one driver can hardly handle the entire audible frequency range limited by the mechanical feature of the driver. In the lower-end speaker market, like what in the phone, where usually only one driver is used, manufacture is capable to control the quality of their product in only a narrow frequency range, while quality outside the important frequency range is less concerned for some reasons.

Firstly, the important frequency range covers most of human's sensitive frequency range, while we are not sensitive to the left frequency range, which leads the quality control outside the main frequency range less meaningful.

Secondly, compensating the quality costs a lot, which increases the overall costs and decrease the competitiveness of the manufacture in terms of price. For example, adding an independent high frequency driver enhances the quality sharply, however increases the cost multiple times. So phones in the market are often equipped with only single speaker driver.

As the result, manufacture control the sensitive range quality and let alone the insensitive frequency range.

\begin{figure}[h]
    \centering
    \includegraphics[width=0.45\textwidth]{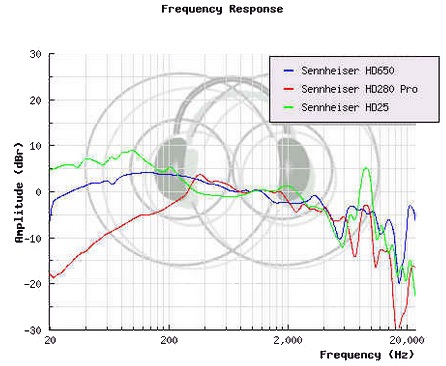}
    \caption{Frequency response of 3 speakers.}
    \label{fr}
\end{figure}

Frequency response presents the quality of a speaker from the perspective of frequency, which weight the quality of a speaker by reflecting the gain or attenuation the speaker provide at each frequency point. It is easy to conduct that the more the response curve is flat, the better voice quality will it provide. Figure~\ref{fr}, captured from the Internet~\cite{3speaker}, presents the frequency response of three speakers which shows that: at low frequency segment, they have similar response curves, meanwhile, at high frequency segment, their response curves are different to each other dramatically. Not only the variances between different models of speakers but also the differences between what of the same model are huge.

Both theoretical analysis and experimental result, which will be shown in the evaluation section, drives us to decide to use the high frequency range response feature, as it carries high variations between each speaker individuals.

\subsubsection{Be Robust with Controlled Stimulus Patterns}

The sampling data collected by many previous work are just the results of uncontrolled input stimulus. For example, in~\cite{dey2014accelprint}, the sampled accelerometer readings are stimulated by random user movement. In~\cite{das2014fingerprinting}, even though the music played could controlled, but the frequency component combinations and variations are determined by the stimulation as well as the abundant noise permutated in the environment. Due to the non-linear features of speakers, like inter-modulations~\cite{brown2006linear}, the recorded sound may contain lots of noises that would make the result unstable. The software-based method like browser configuration tracking also suffers from noise that is user's configuration modification.

\begin{figure}[h]
\centering
\epsfig{file=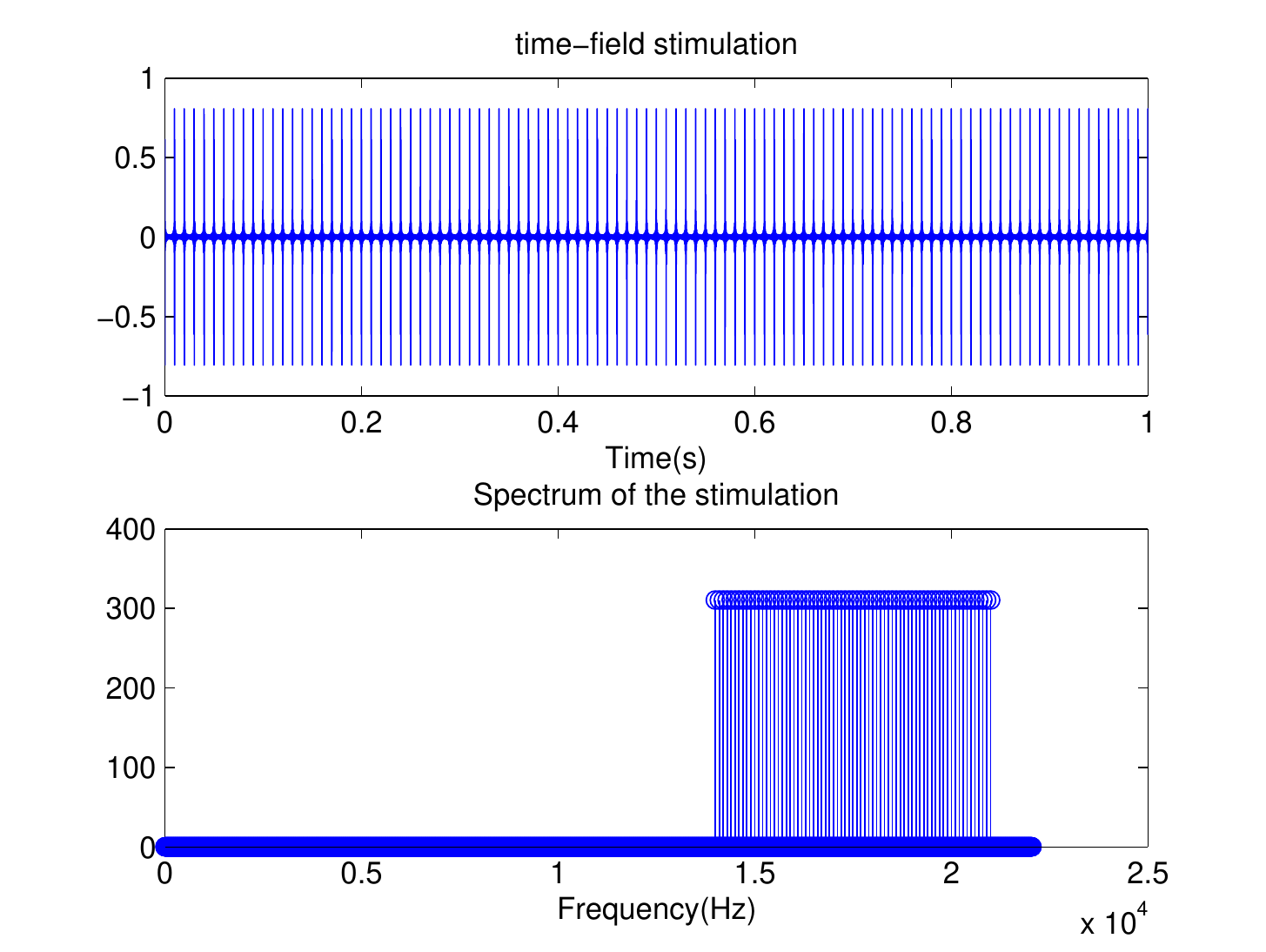,width=0.5\textwidth}
\caption{Stimulation.}
\label{stimu}
\end{figure}

In contrast, we propose to use a controlled audio wave pattern to drive the speaker, so that the results is more robust to random and non-linear factors, and less vulnerable to noises. One example pattern is shown in Figure~\ref{stimu}.

The stimulation lies in a frequency range that interfered only little by the environment. As the spectrum of noise in different environment will be shown in Figure.~\ref{noise}, we found the silent environment in high frequency band provides a perfect test bed for measuring the frequency response of the speaker. It is just the less-interfered environment, controlled stimulation that brought robustness to the scheme.

\section{Design}
\label{sec:impl}

In this section, we introduce how the scheme generate stimulation, calculate frequency response and search the feature in the database.
\subsection{Stimulation Generation}
In our scheme, the android phone itself generates appropriate acoustic signal by playing a period of synthetic sound as the stimulation and itself collects the response from the microphone. Comparing with the passive generation, where the response is highly affected by the stimulation provider, active one, in fact, provide plain, pure and noise-less response resulted from a self-controlled stimulation.

%As the stimulation generation, different from a piece of randomly chosen music which generally contains a wide spread of frequency components and can be easily interfered by noises. What is more, in music, the frequency combinations generally change very quickly from $t_1$ to $(t_1 + \Delta t)$, and the same from $(t_1 + \Delta t)$ to $(t_1 + 2\Delta t)$, which means the speaker output at time  $(t_1 + 2\Delta t)$ will not only depends on current frequency combination, but also the states of previous time. After all, speakers perform mechanical movements to produce sound, and we can have a simple thought experiment to understand such non-linear feature: if we output a 1~kHz sound signal to speaker for one second and then stop the signal immediately, can the speaker stop viberation immediately? No, it will keep viberating for some time. Now, if we output another 300~Hz signal immediately after 1~kHz, can the speaker viberate 300~Hz immediately? The answer is still NO, becasue previous 1~kHz signal still has some impacts on the speaker's viberation.

We didn't use a wave with continuous flat frequency band because the power of the signal is constrained resulted from a very high PAPR (Peak to Average Power Ratio) in that case. We also didn't adopt a frequency shifted music, because the complex frequency combinations make the output unstable because of the non-linearity attribute of the speaker. Instead, we adopted the stimulation shown in figure.~\ref{stimu}. It is consist of a series of cosine wave from 14 kHz to 21 kHz with 100 Hz gap between neighbor frequency points. In order to play the high frequency sound, we set the sample rate of the PCM format input to the android API at 44100 Hz.

\subsection{Feature Generation}
The feature is generated by measuring the frequency response curve. This section illustrate how to generate a feature.

To get the frequency response, we use the spectrum of recording divided by the input. The spectrum of recording is calculated by the FFT. The process of dividing by the input can be neglected since the magnitude in each frequency point is a constant and the response will be normalized later.

In this way, the measured response, in fact, is the response of the whole acoustic chain. However because cascaded system can be regarded as a single system which possesses a system function that is the product of all the subsystem. The signal recorded divided by the signal of the input tells the function of all the system cascaded lying in the chain. And the system functions of media, microphone, chip contribute little variance, which result it to be regarded as flat systems that amplify or diminish the signal evenly to every frequency point in the range. Therefore, the whole system function can be regarded as the amplified or diminished version of the speaker's. As the result, we just use the system function of the whole system function to represent what of the speaker's.

Considering the response feature, in the effective points, the frequency response is signal mixed with interference of noise brought by the environment. While in the gap between effective points, the response is meaningless because there exists only noise. So only the effective points are counted when producing the feature. Besides, in each point, the phase can be neglected because it is easily interfered by the environment. So we only calculate aptitude instead of considering the complex number.

To save communication bandwidth and storage, in this scheme, only aptitude of 71 effective frequency points were counted. And it is not the truth that the more points are sampled, the higher entropy will it accumulate, because the power of the stimulation will be allocated to each frequency point, where not sufficient power leads to not sufficient SNR (signal noise ratio) and a not stable curve thereby. So we use a vector containing the 71 effective aptitude as the representation of the feature.

\subsection{Feature Matching}
Matching two device is just matching the two curve, hence, the vectors they owned. To judge if the two vectors come from the same device, the proof is their similarity. The more similar the two vectors are, the more possible that they come from the same device. Mathematically, the distance between two vectors can be utilized to weight the similarity of two vectors. The shorter the distance is, the more similar they will be. Once the newly received feature is close enough to some existed feature in the database, they will be judged as produced by the same device. Otherwise, a new profile will be set up for the new comer.

In the experiment phase, we just use the brute force algorithm to get the most similar feature vector met before and judge if the distance between them reached a predefined threshold(an experimental value 0.7 is set in the experiment phase) to tell if it is a new user or it is just the user the most similar vector represent. Because ultra-large scale data has not been collected and searched, this scheme runs pretty fast. In fact, with the expansion of the scale of the data, matching users one by one becomes time wasting and not feasible. But this never mask the fact that the float vector can be easily fuzzy searched using Locality Sensitive Hashing or k-NN algorithm. In that case, the searching time complexity can be reduced to nearly a constant.~\cite{gionis1999similarity,yianilos1993data,slaney2008locality}

\section{Evaluation}
\label{sec:eval}
As a practical and feasible fingerprint, the scheme should be inspected in some aspects. For example, fingerprint should be stable as it changes little from time to time, which remind us to check the stability of the frequency response. This section shows our test result to answer the following questions:

\begin{itemize}
\item{\emph{Performance} Can the scheme be applied to large scale user tracking? Specifically, can a large amount of users be distinguished from each other.}
\item{\emph{Stability} How stable is the response curve? Is it feasible for long term user tracking.}
\item{\emph{Interference} How the noise in different environment interfere the performance of the scheme.}
\end{itemize}

\subsection{Experiment Setting}
\begin{figure}[h]
    \centering
    \includegraphics[width=0.5\textwidth]{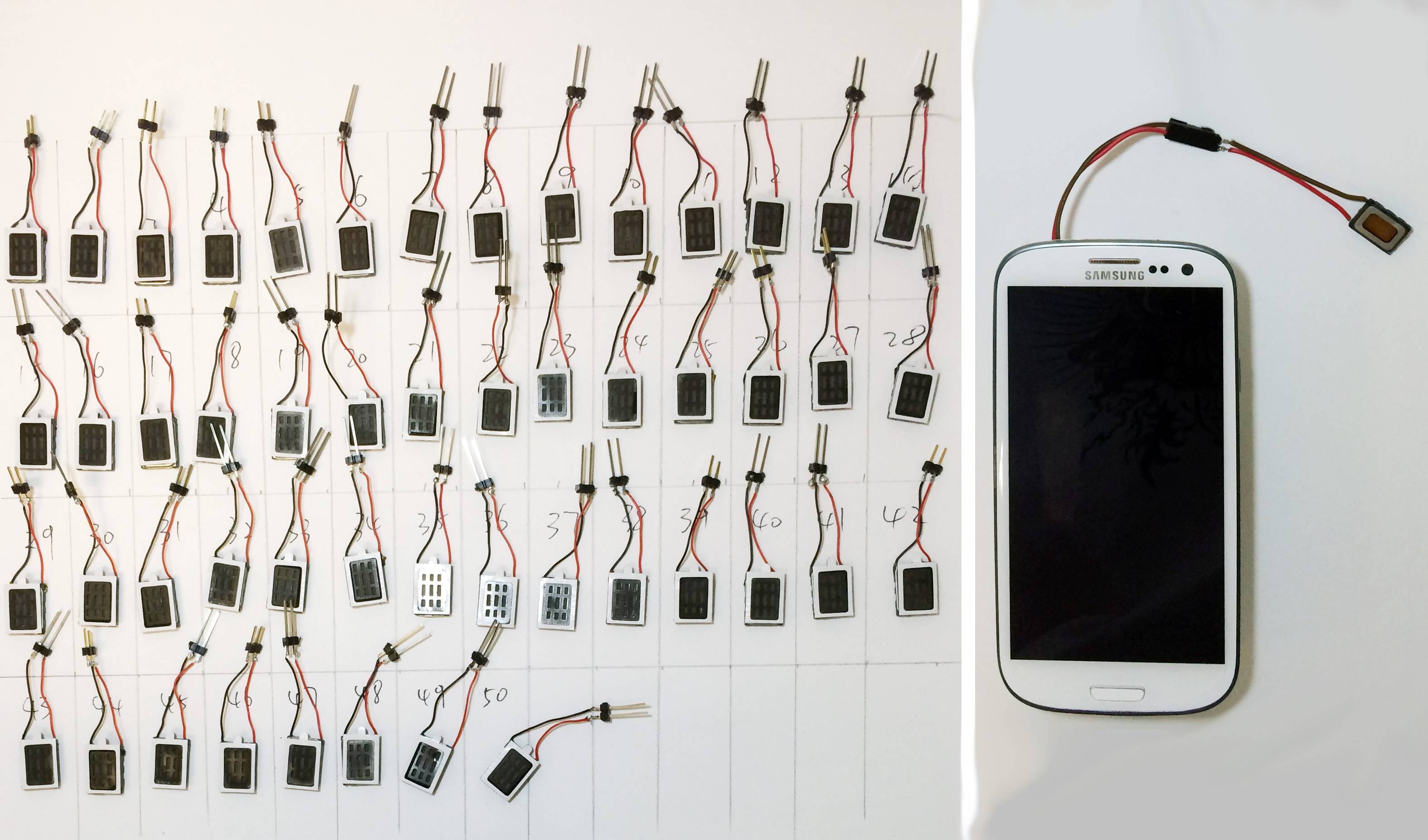}
    \caption{Experiment Equipment.}
    \label{exp}
\end{figure}

According to previous study, it is much easier to distinguish phones of different models and from different manufacturers~\cite{das2014fingerprinting}, so in this paper, we focus on the testing of different phones from the same model. To complete the test economically, we decided to conduct the experiments on 50 OEM speakers on a single Samsung Galaxy S3. We modified the Galaxy S3 by converting the soldered speaker interface into a pluggable socket, as shown in Figure~\ref{exp}, then we purchased 50 OEM (Original Equipment Manufacturer) speakers which are coming from the same assembly line and have continuous Serial Numbers. These speakers are soldered with two-pin plugs, so that they can be easily connected to the phone.

To study the scheme justifiedly, the experiment in conducted in the normal office environment with normal noise level if it is not postulated intentionally in the following part. During the experiment, classmates' conversation is not prohibited on one side to emulate common case, on another to prove the low noise level in the high frequency range.

To each emulated phone, 60 sets of response feature was collected for further evaluation. Thus, totally 3000 vectors have been collected.

\subsection{Metrics}
The metrics listed are used to evaluate the scheme:
\begin{itemize}
  \item{\emph{Feature Distance} Since the feature is actually a vector in N-space, we simply define the feature distance as the Euclidian distance in N-space listed below:
    $$ d(p, q) = \sqrt{\sum_{i=0}^{N}{(q_i - p_i)^2}}$$
    where $p$ and $q$ are two feature vectors defined as: $$p=(p_0, p_1, \cdots, p_{N-1}), q=(q_0, q_1, \cdots, q_{N-1})$$}
\item{\emph{Similarity} We use similarty to measure how likely the two features $p, q$ are coming from the same phone, and it is defined as $$ 1 - d(p,q)$$}
%\item{\emph{False Positive} The case, where feature vector produced not by the phone, no matter if it is in the data base, is judged being produced by the phone, constitutes the false positive.}
\item{\emph{False Positive} We define a case as false positive if phone A is falsely recognized as another phone B based on the input features.}
%\item{\emph{False Negative} The case, where feature vector produced by the phone whose feature was already in the database is judged as an alien device, constitutes the false negative.}
\item{\emph{False Negative} We define a case as false Negative if no matches can be found in the database for features from phone A that actually does exist in the database.}
\item{\emph{Entropy} The logarithm of (the size of the distinguishable set) to base 2 is the entropy of the scheme. The distinguishable set is the set that all the contained element can be distinguished from each other by the feature produced.}
\end{itemize}

\subsection{Performance}
At first, we planned to count the quantity of false. So, the 3000 feature vectors are input to the process in a random sort. The output is checked with right answer to count false positive and false negative. We are very happy to tell that no false positive nor false negative found in the 3000 test cases. But it can hardly justify the performance of the scheme when the quantity of test cases increases sharply. So, we refer to the distribution of the similarity to calculate the performance in the large scale case.
\subsubsection{Distribution of similarity}

\begin{figure}[ht]
    \centering
    \epsfig{file=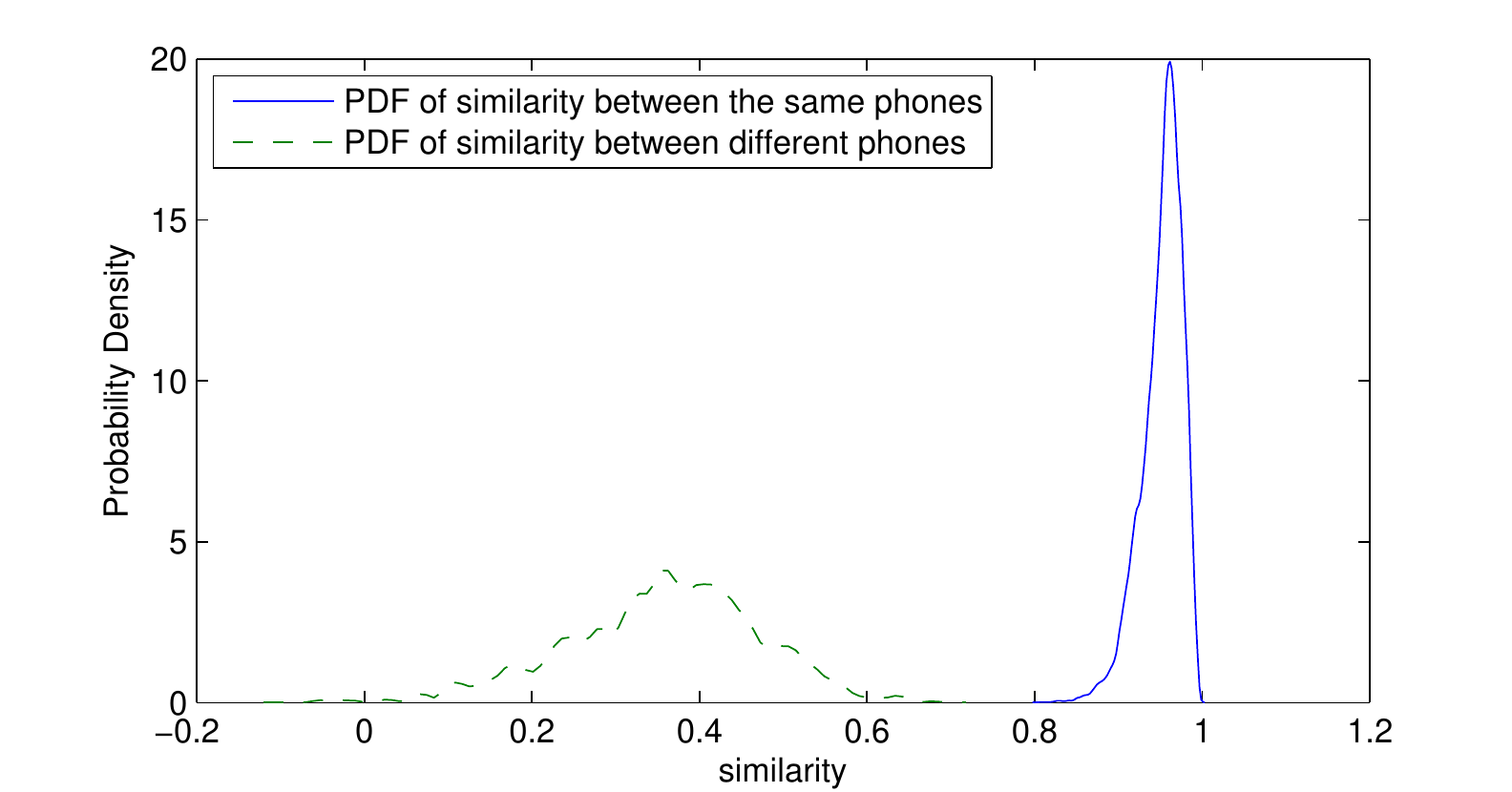,width=0.5\textwidth}
    \caption{Distribution of Similarities.}
    \label{dis}
\end{figure}

We found there is a gap between similarities of the same phone and the similarities of different phones, which is the main reason of the good performance. We investigated the distribution of similarities between different phones ($sim_{corr}$) and within the same phones ($sim_{self}$) respectively. Specifically, in terms of $sim_{self}$, to each device, comparison between the 60 features results to $C_{60}^{2}$ $sim_{self}$. So, totally 50* $C_{60}^{2}$ $sim_{self}$ are collected. In terms of $sim_{corr}$, there are $C_{50}^{2}$ devices pairs, where 60*60 similarities can be calculated in each pair. So, totally, $3600*C_{50}^{2}$ $sim_{corr}$ are collected. The PDF (probability density function) of the distribution is shown in Figure~~\ref{dis}.

The gap between the PDF of $sim_{corr}$ and $sim_{self}$ revealed the reason that we found no false. Specifically, the similarity between different phones spans in a range which has no common part with what of similarity between the same phones. Generally speaking, the maximum value of the $sim_{corr}$ is less than the minimum value of $sim_{self}$. So, facing a newly arrived feature vector, the similarity between it and its' nearest neighbor is calculated. It can be concluded that they comes from the same device if only this similarity locates at the right side of the gap. Otherwise, the feature comes from a unknown device.

Because the error rate of the scheme is directly linked with the probability distribution over the gap. However, under this setting, the probability of feature's crossing the gap is unknown resulted from lacking with such observation. So, we shift to get an analytical description of the PDF.

\subsubsection{Distribution Fitness}
\begin{figure}[ht]
    \centering
    \epsfig{file=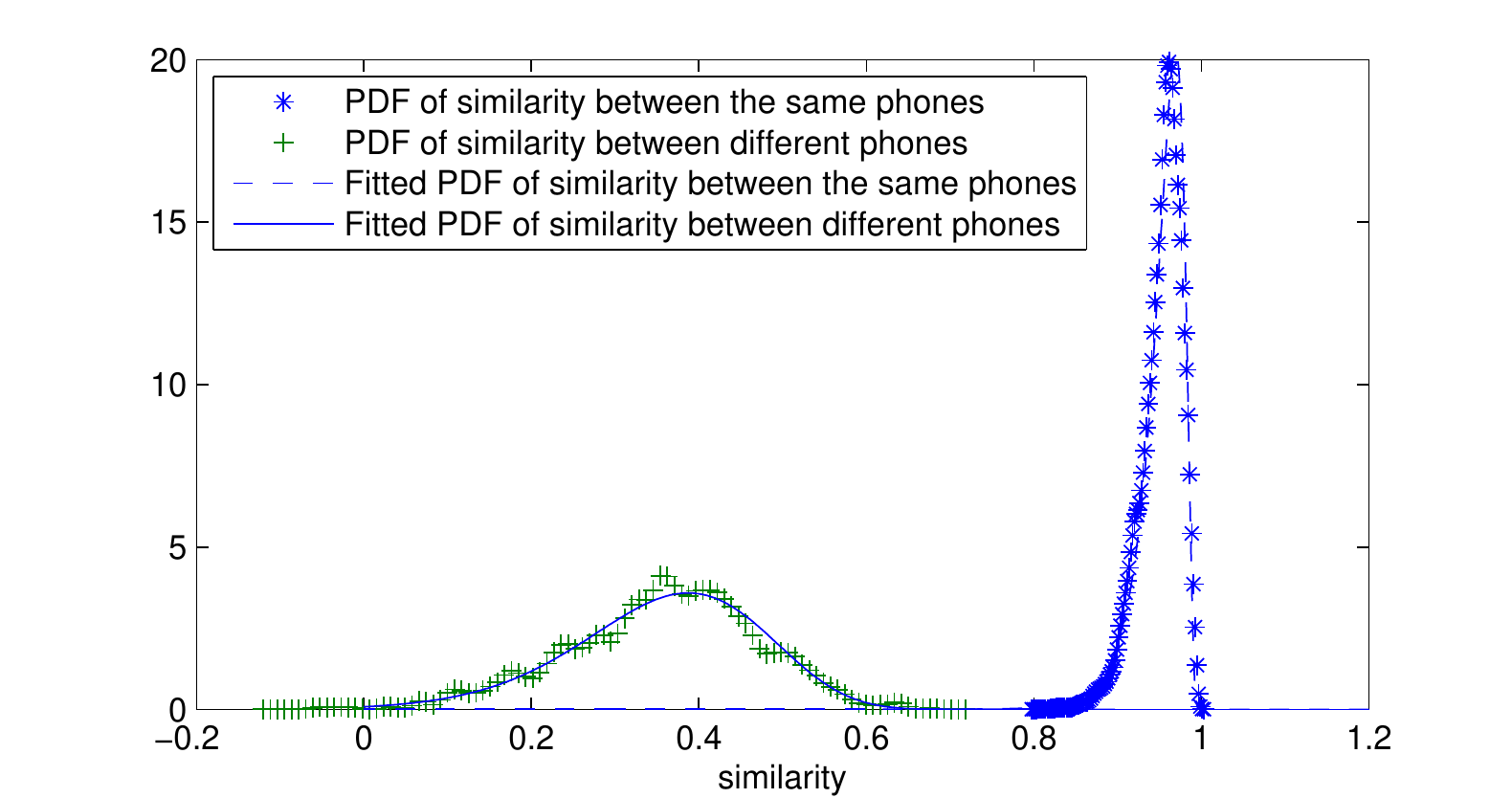,width=0.5\textwidth}
    \caption{Fitted Distribution of Similarities.}
    \label{fit}
\end{figure}

We inspected the two distribution to find their proper distribution type respectively. We found that both of them are unsymmetrical shaped. So we traversed all the common seen distribution to find the type that fit the observations well. We tried to fit the data to 20 continuous distribution types. After analyzing the fitness, we found that the 2 types of distance derived from observations to feature vectors from either the same phones or different phone pairs fall into Lognormal Distribution well. The fitted distribution is shown in Figure.~\ref{fit}.

Because distance falls in Lognormal distribution. The similarity, which is 1 - distance, falls in the distribution with the following PDF:

\begin{align}
f_{self} = \frac{1}{(1-sim_{self})\sigma\sqrt{2\pi}}e^{-(\ln(1-sim_{self})-\mu)^{2}/2\sigma^{2}} \notag
\end{align}

Where the fitted parameter gives $\mu = -3.17698, \sigma = 0.546804$.
\begin{align}
f_{corr} = \frac{1}{(1-sim_{corr})\sigma\sqrt{2\pi}}e^{-(\ln(1-sim_{corr})-\mu)^{2}/2\sigma^{2}} \notag
\end{align}

Where the fitted parameter gives $\mu = -0.457726, \sigma = 0.178714$.

\subsubsection{Scale}
\begin{figure}[ht]
    \centering
    \epsfig{file=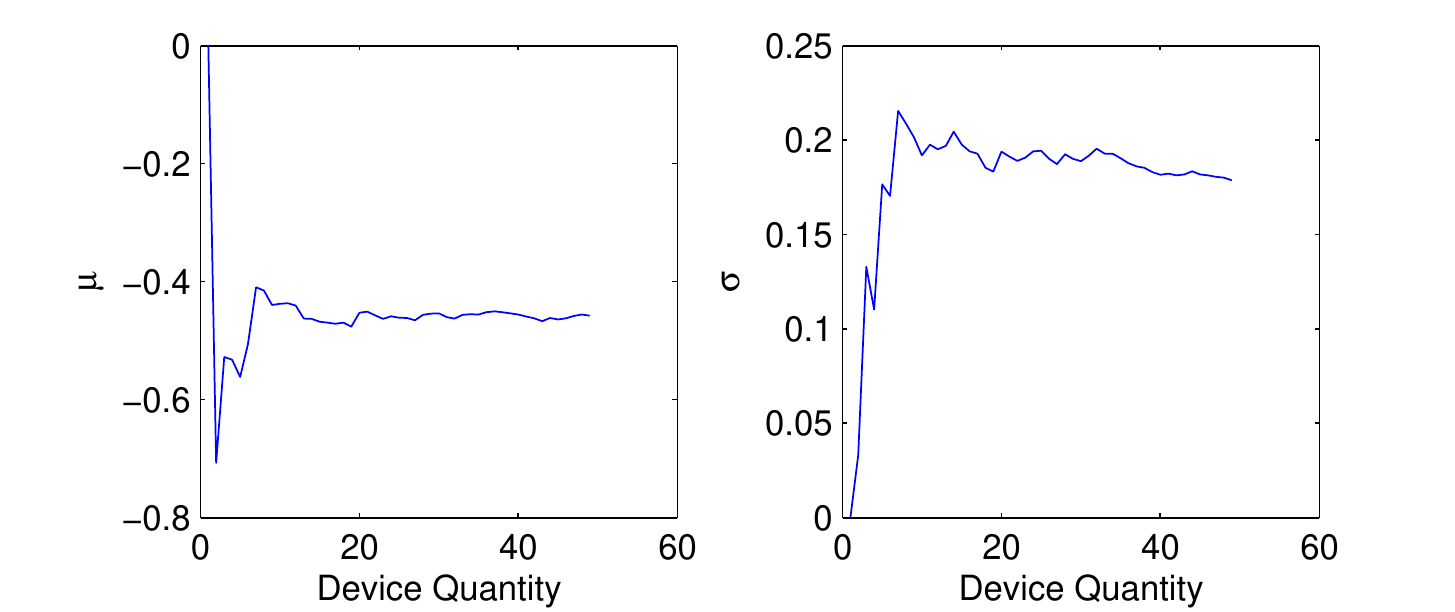,width=0.5\textwidth}
    \caption{Parameter vs Device Quantity.}
    \label{scale}
\end{figure}

We prove that the distribution can be applied to the large scale case. Doubt may be casted on the assumption that the distribution may be correlated with the quantity of the phones. We argue that the distribution of $sim_{corr}$ changes little with the increasing of device quantity, which implies that the error rate of the scheme doesn't increase when the quantity of the devices increases. Changes of parameters $\mu$ and $\sigma$ according different quantity of devices are shown in Figure.~\ref{scale}.

As we can see, the parameters converge to constants when the quantity increases. Based on the result, we assume the model is suited for large scale similarity representation.
\subsubsection{Error Rate Analysis}
We give the theoretical analysis to the error rate based on the model deducted from the prior part. We analyze the false positive rate and false negative respectively first. We then calculate their sum and analyze the error rate under multiple sampling time case. At last, we tell the scheme operator that the parameter can be tuned such that it performs best satisfying the cookie substitution case.
\begin{figure*}[ht]

%\begin{minipage}[ht]{0.5\textwidth}
%    \centering
%    \epsfig{file=./figure/eval/fp.eps,width=\textwidth}
%    \caption{False Positive Rate vs Similarity ($\alpha$).}
%    \label{fp}
%\end{minipage}\begin{minipage}[ht]{0.5\textwidth}
%    \centering
%    \epsfig{file=./figure/eval/fn.eps,width=\textwidth}
%    \caption{False Negative Rate vs Similarity ($\alpha$).}
%    \label{fn}
%\end{minipage}

\begin{minipage}[ht]{0.5\textwidth}
    \centering
    \epsfig{file=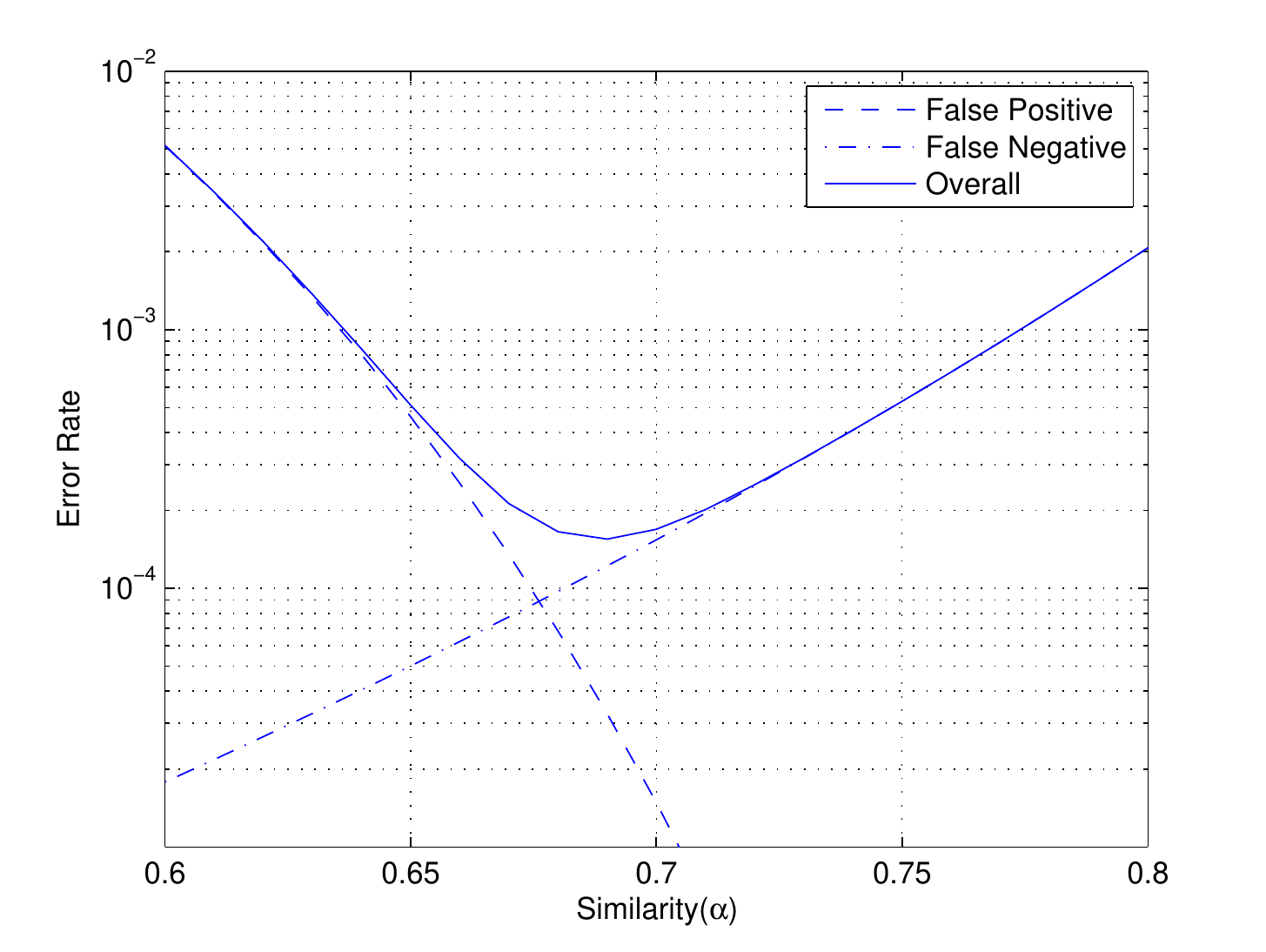,width=\textwidth}
    \caption{Error Rate vs Similarity ($\alpha$).}
    \label{er}
\end{minipage}\begin{minipage}[ht]{0.5\textwidth}
    \centering
    \epsfig{file=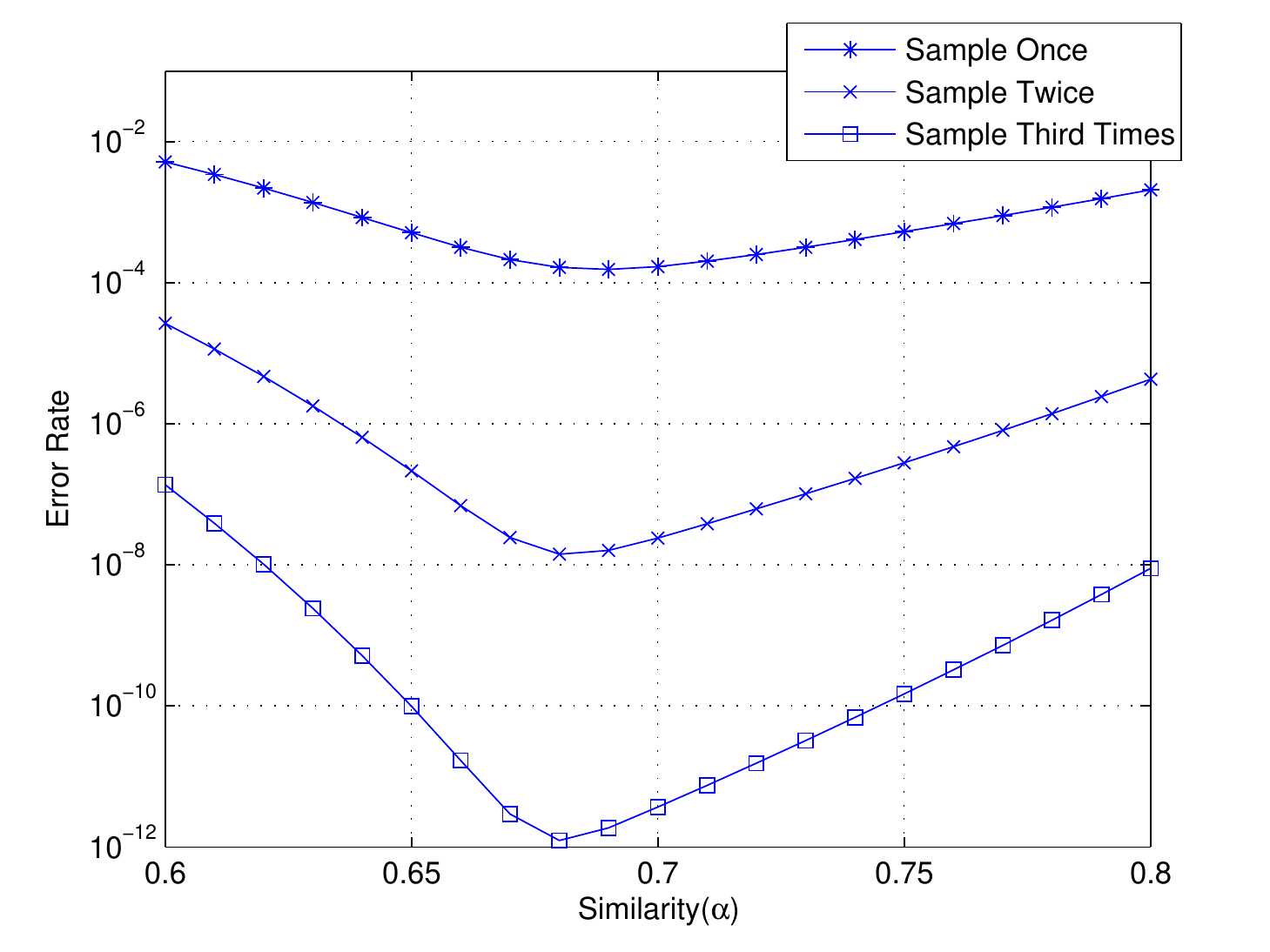,width=\textwidth}
    \caption{Error Rate vs Similarity ($\alpha$).}
    \label{twice}
\end{minipage}
\end{figure*}

\textbf{False Positive}
Theoretically, if an alien observation to $sim_{corr}$ crossed the gap and fell into the range occupied by the  $sim_{self}$, it may be regarded as produced by some device already in the database. The probability of this case is $1-F_{corr}(\alpha)$, where $\alpha$ is the threshold set by server. Curve false positive in Figure.~\ref{er} shows the relationship between $\alpha$ and error rate.

There exists another case, which also leads to false positive. Feature vector produced by Alice may have a $sim_{corr}$ with Bob's that is higher than $sim_{self}$ of Alice her self's, which lead server to output Bob. The probability of this case is $\int_{\alpha}^{1}f_{corr}(x)F_{self}(x)dx$, which is preeminently less than $1-F_{corr}(\alpha)$. As the result, it is neglected when calculating the error rate.

\textbf{False Negative}
An observation to $sim_{self}$ may fall into the range belongs to $sim_{corr}$, which misleads the server to output null instead of the right answer. The probability of this case is $F_{self}(\alpha)$, as it is shown in Curve False Negative of Figure.~\ref{er}.

\textbf{Overall Error Rate}
The error of the scheme is defined by the sum of false positive and false negative. The error rate calculated by the sum of the two kinds of error rate thereby. It changes according to $\alpha$, which is shown in Figure~\ref{er}. The figure tells that lower $\alpha$ brings to more false positive while higher $\alpha$ leads to more false negative. The valley point of the curve locates at 0.69, which imply that setting threshold to 0.69 gives the the best performance.

As we can see, the error rate is around 1.55*$10^{-4}$, when the threshold is set at 0.69.

\textbf{Performance Enhancement}
Sampling multiple times elevate the performance sharply. Collecting each feature vector cost only little. And each sampling can be regarded as independent, which therefore inspired us to collect feature more than once to decrease the error rate. For example, if we collect 2 samples each time, the error rate decreases sharply because the false positive happens only if both two samples are false positive sample, and the false negative happens only if both two samples are false negative sample. Figure.~\ref{twice} shows that the error rate of the twice scheme is around 1.41*$10^{-8}$, when the threshold is set at 0.68. Hence, 1.23*$10^{-12}$ error rate can be achieved if 3 times sampling is adopted.

\textbf{Biased Case} The threshold parameter can be tuned to satisfy different cases. For instance, as the substitution to cookie, the consequence brought by false positive and false negative is not equal. Specifically, clearing cookies often results to regard an old user as a new comer, which is similar to the false negative. While seldom will a piece of cookie be judged wrongly as someone other's with neither rhyme nor reason, that is false positive. As the result, the tolerance to false negative of the server used present is much higher than what of false positive. To this end, the threshold $\alpha$ can be elevated to trade false negative performance for the performance of false positive.

\subsection{Stability}
\begin{figure}[ht]
    \centering
    \epsfig{file=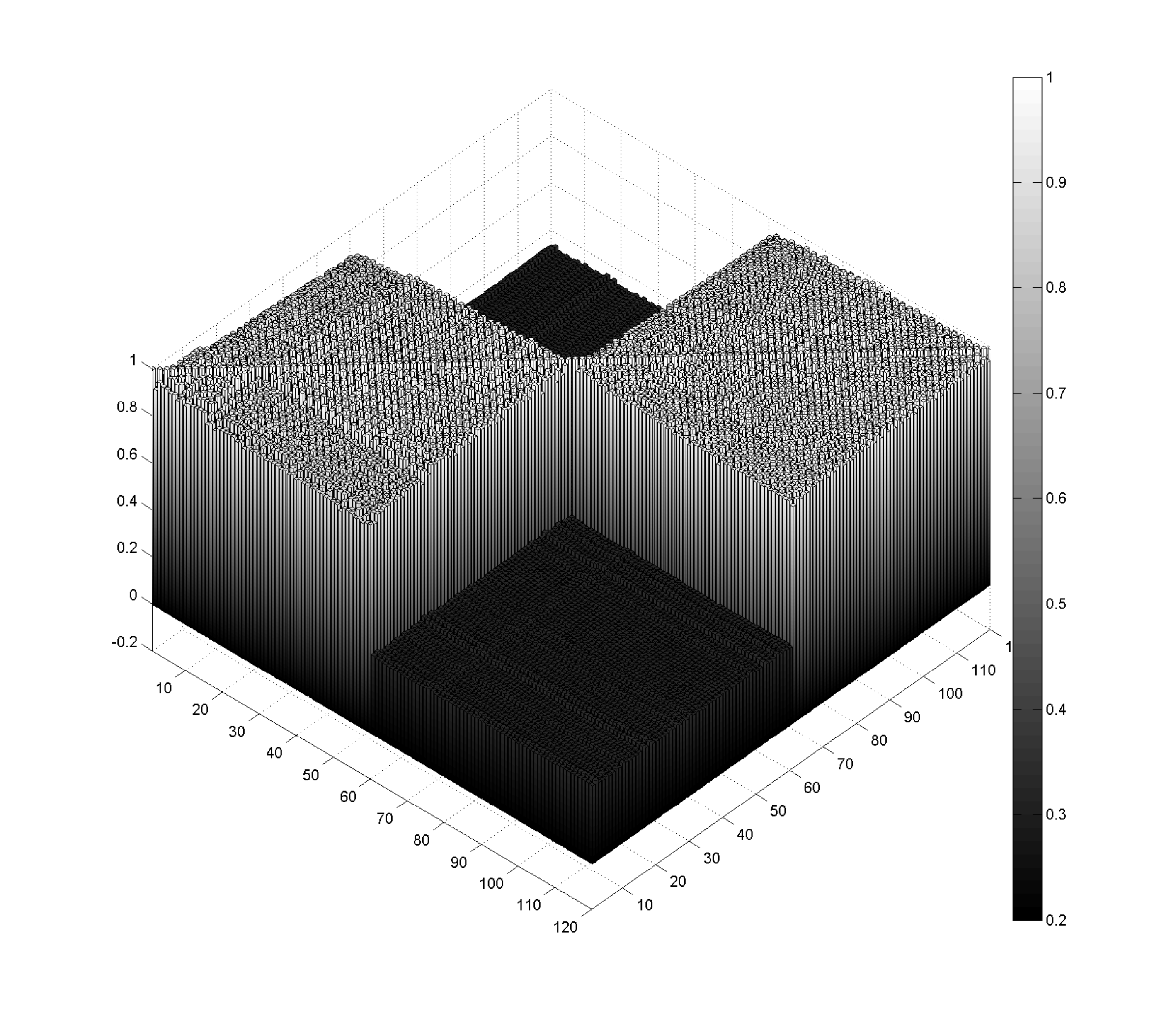,width=0.5\textwidth}
    \caption{ Correlated Similarity.}
    \label{stable}
\end{figure}

We argue that the higher frequency response feature is a kind of long-term stable and unchangeable feature. To be a kind of identity, the feature should be stable span a range of time. But in the case of cookie, different people clear their cookie with different time gap. Some people never clear their cookie while some others never save the cookie, which casts doubt on the stability of the cookie as a kind of identity. To prove our scheme's stability, we chose 2 speakers randomly and collect feature vector every 1 hour to each emulated phone. As the result, we have collected 60 feature vectors to each phone totally. The vectors produced by the first phone are labeled from 1 to 60, while the vectors produced by the second one are labeled form 61 to 120. Figure.~\ref{stable} shows the similarity between the 120 vectors.

As concluded from Figure.~\ref{stable}, there is no obvious decreases in similarity between feature vectors within the same phone collected from the first hour to the last hour. Also, we haven't observed obvious increase in similarity between the two phones from hour to hour. So, the experiment concludes that the higher frequency response feature is long-term stable.

\subsection{Interference}

\begin{figure*}[ht]
    \centering
    \epsfig{file=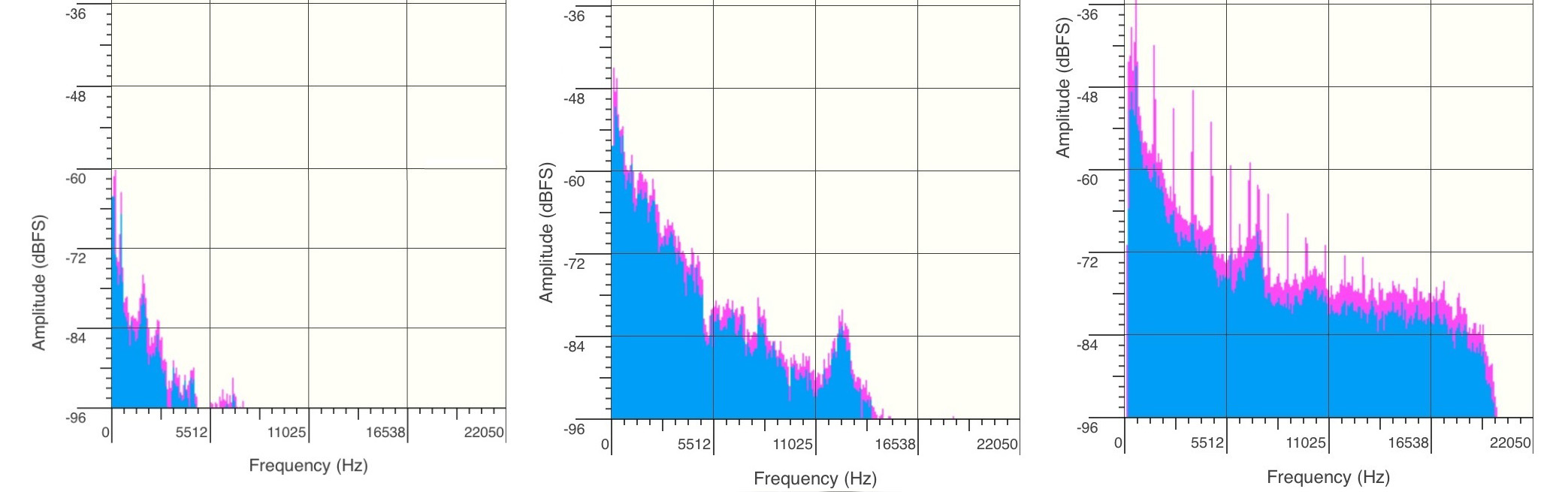,width=\textwidth}
    \caption{Noise in Office, Street and Metro.}
    \label{noise}
\end{figure*}

The higher frequency response is affected by the noise in the higher frequency range, which is pure and silent in most cases. In order to prove the ability to anti interference, we have tested the scheme in different environment with different noise, ranging from office, street, metro station. The result is positive in all cases except the metro station. Figure.~\ref{noise} shows the spectrum of noise on the air in the 3 environment. We will present both qualitative and quantitative analysis to the anti-interference ability of the scheme.

\textbf{Qualitative Analysis} In the effective frequency range, thus from 14kHz to 21kHz, The environment is silent in the case of office and street, though there is loud human being's voice and other noise, which doesn't locate at the effective band. So the response feature can be calculated with only little interference. In the case of metro, the noise spans all the sampling frequency range includes the effective band, which overwhelm the signal broadcasted and make the calculation result meaningless. So, we conclude that the scheme works if only the high frequency band is silent.

\begin{figure}[ht]
    \centering
    \epsfig{file=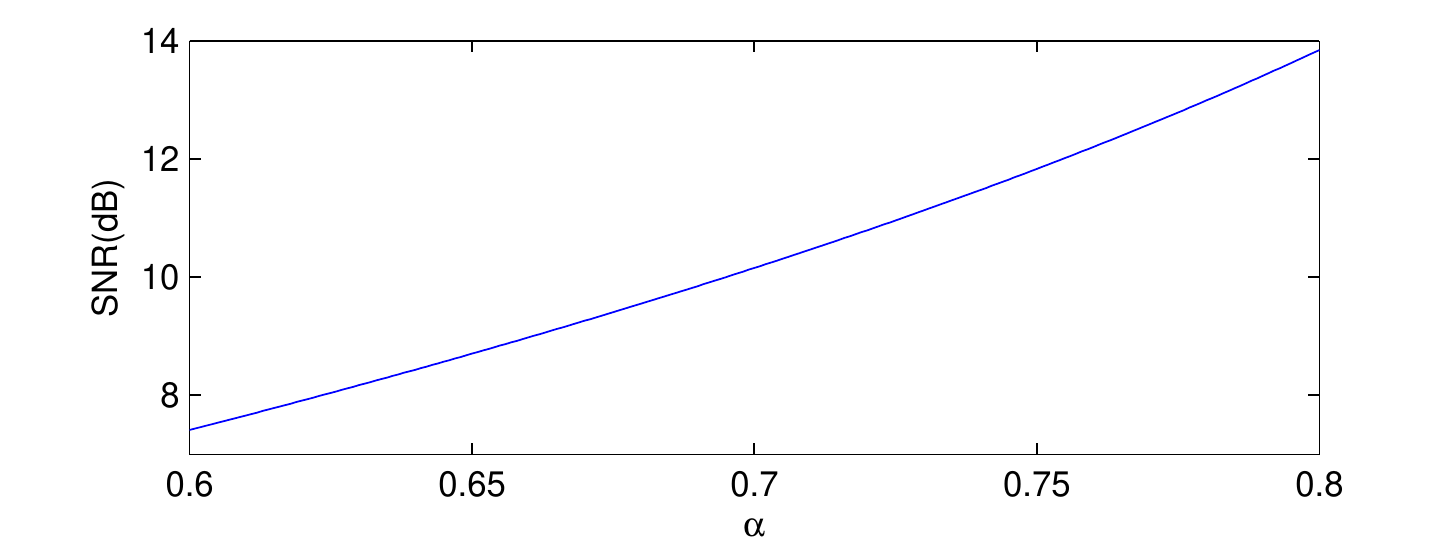,width=0.5\textwidth}
    \caption{SNR requirement over different $\alpha$.}
    \label{snr}
\end{figure}

\textbf{Quantitative Analysis}
In this section, we try to find out the highest noise level at which the scheme works. To simplify the problem, we reasonably assume that the feature is absolutely stable and all the distance between the features of the same phone is resulted from the interferences in the environment. The sampled spectrum of signal is denoted as $\vec{X}$, while the noise is denoted as $\vec{N}$. We also assume that there is little correlation between $\vec{X}$ and $\vec{N}$, and the expected mean of the $\vec{N}$ is zero (We assume like this because the noise is often white noise), which leads $\vec{X}$ and $\vec{N}$ to be regarded orthogonal and $\vec{X}\cdot\vec{N} = 0$ thereby. The Similarity calculated in fact is:

\begin{align*}
   &1-\sqrt{(\frac{\vec{X}}{|\vec{X}|}-\frac{\vec{X}+\vec{N}}{|\vec{X}+\vec{N}|})^2}
 \\&=1-\sqrt{2-2\frac{\vec{X}(\vec{X}+\vec{N})}{|\vec{X}||\vec{X}+\vec{N}|}}
 \\&=1-\sqrt{2-2\frac{|\vec{X}|}{|\vec{X}+\vec{N}|}}
\end{align*}

We consider the false negative in the interfered environment while neglect the case of false positive, because noise can easily make a feature distorted, but hardly make a feature similar to another. The server output right answer when this similarity between the 2 feature vectors is higher than a threshold $\alpha$. Thus:
\begin{align*}
&1-\sqrt{2-2\frac{|\vec{X}|}{|\vec{X}+\vec{N}|}} > \alpha
\\&\Rightarrow\frac{|\vec{X}|}{|\vec{X}+\vec{N}|} > \frac{1+2\alpha-\alpha^2}{2}
\\&\Rightarrow\frac{|\vec{X}|^2}{|\vec{X}|^2+|\vec{N}|^2} >\frac{\alpha^4-4\alpha^3+2\alpha^2+4\alpha+1}{4}
\\&\Rightarrow\frac{SNR}{SNR+1} > \frac{\alpha^4-4\alpha^3+2\alpha^2+4\alpha+1}{4}
\\&\Rightarrow SNR > \frac{1+4\alpha+2\alpha^2-4\alpha^3+\alpha^4}{3-4\alpha-2\alpha^2+4\alpha^3-\alpha^4}
\\&\text{Where SNR is }\frac{|\vec{X}|^2}{|\vec{N}|^2}
\end{align*}

The SNR can be calculated in this way according to Parseval's theorem, which indicate the power of a signal can also be the sum of the power of each frequency component. While the power of each component is the square of its amplitude. Therefore, the square to the normal of the vector is just the power.

The relationship between $\alpha$ and error rate, thus Figure.~\ref{snr}, shows the SNR requirement in avoiding false positive in different $\alpha$ setting. As we can see, in the normal setting, thus $\alpha = 0.7$, the SNR requirement is 10 dB. That means the scheme output right answer if only the SNR in the effective frequency band in higher than 10 dB. Don't forget that the noise power is only counted for those locate at the effective frequency points, which possesses only little of the overall noise power.
\subsection{Entropy}
We calculate entropy in this part, because entropy is important in evaluating an identity scheme. Entropy weights how many information the identity carries, hence how many devices can be distinguished from each other. Specifically, in order to distinguish a set of devices whose size is N, at least $log_2N$ bits entropy should be carried during a round of identification. Therefore, we analyze how many devices can be distinguished by deploying our scheme.

After setting the threshold parameter $\alpha$ to the optimized one, the error rate can be calculated accordingly. So, the entropy can be calculated if only the relationship between error rate and the size of the distinguishable size is decided. Approximately $1/error\_rate$ devices can be distinguished at the given $error\_rate$, because less than one error will be found expectedly. As the result, we regard all the $1/error\_rate$ devices distinguishable accordingly. The entropy the identity carries is $-log_2(error\_rate)$ under the settings thereby.

We believe that each feature transferred back to the server carries entropy. So, with the increasing of feature vectors used linearly, the error rate decreases geometrically and the entropy increases linearly, because of the independence between 2 samplings. As calculated before, the error rate is at 1.55*$10^{-4}$, if one feature vector is utilized to make judgement. According to the error rate, 12.6 bits entropy can be achieved in the single sampling case. Hence,we can get 26bits in the double sampling case or 39.6 bits in 3 times scheme.

\section{Application}
\label{sec:application}

The device ID extract from our proposed scheme can not only replace traditional cookies, but also be used beyond that, and this section will give brief introduction to some of them.

\subsection{Stolen phone tracing and self-destruction}

Recently, lawmakers in California has approved a so-called ``kill switch'' bill, which requires all smart phones sold in the state to have anti-theft software installed, so that once the device is lost or stolen, it cannot be used any more, even after a hard reset~\cite{killswitch}.

However, in order to achieve that goal, the first step is being able to uniquely identify a device. This is not a trivial task, given the fact that the phone could be reset, re-flashed with different operating system image, or even modify the IMEI code via software. In other words, since every piece of current device information is stored in Flash memory, and the Flash memory is under the control of adversaries, there is nothing can prevent them to modify such information and defeat the ``kill switch'' mechanism.

Our speaker-based device ID can help address the challenge. Any changes in the software cannot change our hardware based device IDs. So, in order to check if current phone has been reported as stolen, the system vendors need only to perform a quick and un-noticeable test, and then look up the generated device ID in the stolen phone database.

To avoid such detection, the adversaries have to modify the hardware, but the cost is high, not only because the extra money to buy new parts, but also the time and skills to perform such hardware modification (especially when the phones are becoming more difficult to be dissected).

\ignore{
Phone stolen is common seen in the world. At present, the privacy stored in the phone is playing a as equally or more important role comparing to the hardware. A stolen phone may leak privacy such as photo, Message, autofilling account and password etc. Some application is designed to inform the owner the position of the stolen phone. But as they are software-based, it is easy for a thief to wipe out all the data including the application. Besides the data formatting, without network, hardly will this kind of application work.

However in the case of hardware-based frequency response feature identity, the service provider can report the location of the stolen phone to the owner, once the phone's identity is captured by his neighbor (without network requirement) or by the application installed by the thief. In a word, the stable hardware feature makes stolen phone tracking much more easier.

In this setting, stolen phone can be found even the theft phone has no Internet accessing.
}

\subsection{Location information broadcast and relay}

Many applications require getting position information to complete some useful functions. For example, instant message applications can let you know and make friends with people nearby. However, current designs require user to grant the applications to access user's current position, which users often decline, either due to the privacy concerns, or avoid overly power consumption used by GPS subsystem.

But with our proposed scheme, applications can easily share and relay position information, and following is a typical scenario. Suppose there are many people in a conference room, but only one of them turned on the location service, so the server can put information of device ID and the location of that conference room into a database. Now the application will periodically play the specially crafted sound, which can be captured and cross-fingerprinted by other phones nearby. Once the device ID is extracted, those smartphones will query the database on the server, and retrieve the location information generated by another phone with GPS turned on. Once a new phone get its location information, it starts to broadcast its identity, and thus the location information can be relayed across the whole conference room.

\ignore{
The  Application, however, can get around this permission if a near the user broadcast his identity.

Imaging in a church, if only a single prayer opened the GPS, thus his application can read and upload the position information, and this device broadcasts his identity which can be recorded by other prayers nearby, application installed on neighbor prayers can query the server the position of the identity he listened and make it as his own location. Recursively, all the prayers' location can be accessed with only one prayer who granted location permission.

Admittedly, the location information gotten is not as precise as what from GPS. But with no doubt it is the best solution under the condition that no GPS information is accessed.
}

\subsection{Indoor tracking}

Indoor tracking has a huge market potential, with which supermarket and department stores can send coupons and targeted advertisements to their customers. There are already several technologies available, like Bluetooth based iBeacon from Apple~\cite{ibeacon}, and WiFi based solutions~\cite{wifiIndoorTracking}. The device ID proposed in this paper can also be used in this scenario. First, whenever the user enters a supermarket, her phone will receive a signal to trigger the periodically playing of the inaudible sound, which is actually equal to broadcast its device ID from time to time. Such broadcasting will be received by microphones deployed all around the supermarket, then a cross-fingerprinting is performed, and a unique device ID extracted. By correlate the device ID with the microphone location, it is easy to know the route of the user in the supermarket, what her favorite is, and what is still under consideration, etc. With the same technology, it is also possible to associate the purchase history to a specific device ID, simply by putting a cross-fingerprinting microphone near the check-out counter.

\ignore{
Nowadays, many mall related apps aiming at providing the discount information and advertisement published by the mall operator are popular among the consumers. Consumers rarely would like to receive targeted advertisement since that require all the shopping history is recorded by the mall. So, often times we buy something anonymously and don't log our shopping record to the application intentionally because of the privacy and convenience. However, this application can broadcast the ultrasonic, which carries the identity of the device. And the shopping mall can be installed with a lots of recorder distributed all around. As the result, once the customers enter the mall carrying the phone, the routine of the customers can be traced respectively. Besides, the sales record can be correlated with devices by capturing the identity of the device near the checkout counter. In this case, all the app users' shopping record can be extracted and data mining can be invoked to push targeted advertisement to user's phone. More importantly, potential customer who left no shopping record but stay for a while around some products can be also targeted pushed information, which is highly possible to hit the potential customer.
}

\section{Discussion}
\label{sec:discuss}
In this section we will discuss the potential limitations of our proposed method, more specifically, the interference from background noises, and the detection of application doing fingerprinting.

\subsection{Interference from background noises}

Although our proposed scheme has a special design on frequency combinations at about seventy discrete frequency points, it could still fail to extract unique device IDs under environments saturated with high power noise signals, like train station, crowding restaurant, etc.

To overcome such limitation and make our method work even under low Signal-Noise-Ratio (SNR), we may try to use some advanced methods borrowing from communication area, and one example is ``Spread-spectrum Communication''~\cite{spread}.  Spread spectrum communication generally makes use of a sequential noise-like signal structure to spread the normally narrow band information signal over a relatively wide band of frequencies. It can even do frequency-hopping where information is sent following a sequence of pseudo random frequencies. The receiver can reproduce the same pseudo random sequence and thus is able to correlate the received signals to retrieve the transmitted information~\cite{spread}.

Inspired by the spread-spectrum communication, we can modify the scheme accordingly. In each effective frequency point from 14~kHz to 21~kHz, the original mono tone sine wave is modulated with a pseudo random sequence, such that the energy originally in the frequency point spreads to a frequency range, the width of which is decided by the rate of the pseudo random sequence generation. As the result, the distributed energy decreases the energy density sharply while the overall signal energy keeps unchanged, since the consumed bandwidth increases. Later, the recorded audio data will be sent to a band pass filter and de-spread to recover the sine wave. Finally, the recovered sine wave at each frequency point has different amplitude because the speaker attenuates the signals, which reflects the features of that speaker.

%With the spread-spectrum technique, the frequency response curve can be measured in high noise environment. Besides, the usable frequency range can be extended to human's hearable range, because of the sharply decreased energy density.

\subsection{Device ID for smart phones of different models or from different manufacturers}
In this paper, we only evaluated the features of 50 OEM speakers for Samsung Galaxy S3. All the speakers are coming from the same assembly line with continuous Serial Number printed on them. We did not extend our study to smart phones from different manufacturers because of the assumption that speakers from different manufacturers are generally easier to be differentiated, and this has been confirmed by previous work~\cite{das2014fingerprinting}.

Even in the worst case that above assumption fails, we would propose to incorporate other hardware feature or information into the device ID. For example, the CPU type, memory capacity, operating system version, etc. According to previous studies, an app can get all above information without requesting any special permission~\cite{zhou2013identity}.

\subsection{Detection of audio fingerprinting operation}

Although an Android application based our proposed scheme can disguise itself as legitimate one by requesting microphone accessing permission for other legal use, it is still possible to detect if such application is trying to perform audio fingerprinting or not. For example, it is required to do Fast Fourier Transform on recorded response in order to generate audio stimulus, so with some code analysis, it is possible to detect the existence of such suspicious operations though it can be hidden into the equalizer processing as if it is enhancing the audio quality. However, if an application's original function can include these operations, then the detection problem is still very hard.

% Some researchers may argue that the stealthy data transferring may be probated by dynamic or static analysis. However, in fact, both playing and recording part are legal. The playing mixed with ultrasonic can be hidden in the equalizer, which emphasis or attenuate some frequency component to enhance user's listing feeling. The recording process possesses no any differences comparing with the version without tracking. Even by invoking taint analysis, the taint will be cut off between speaker and microphone, because there is no link between data from speaker to microphone.

\
\ignore{

The scheme achieved that a device fingerprint can be generated without limitation of stimulation. The speaker broadcast freely without user's awareness while some other scheme require the stimulation either generated by the user or generated with user's awareness. For example, fingerprint from accelerometer require that the phone should be moving. When the user is sleeping and the phone stays silently, the fingerprint can hardly be generated. Fingerprint from lower frequency response require some audible voice be played, which attracts the user's attention.

The scheme also achieved high entropy. The error rate of the judgement is not correlated with the quantity of the device in total, which imply very high entropy by multiple time sampling. The calculated entropy is sufficient for distinguishing all the phones in the world. But our work still have some drawback in some aspects, as listed as follow:

\textbf{Theoretical Support} The distribution of the similarity is fitted to some well-known distribution type to get ideal and theoretical distribution. But we still can not tell the reason the similarity falls to the lognormal distribution. Thus, we haven't found out the physical implication of the distribution. Similar to the postulation of Planck black-body radiation law, the radiation equation is deduced from the fitness of the statistical data collected from experiment.

\textbf{Recording Blocking} In some applications, recording is inevitable, without which the application may not work. In some others, recording permission is redundant, which may drive the user to cast doubt on the perniciousness of the application. Though we postulated some way to get around the recording permission, it is not guaranteed that application can read recording file with the stimulation as free as with recording permission.

\textbf{Interferences} As the experiment result shows, in some environment with extreme high noise level, such as metro station, the feature vector collected is distorted to an extent that disable the judgement. We can hardy come up with reasonable solution to cope with those environment.

}
\section{Related Work}
\label{sec:rel}
\textbf{Software Fingerprint} In terms of software feature, many browser configuration information can be exploited to differentiate device, such as User Agent, fonts installed, plugin information, benchmark etc~\cite{eckersley2010unique,mowery2011fingerprinting,boda2012user,takeda2012user}. Besides the browser, OS version, Kernel version, application list can all be utilized to distinguish devices. Different implementation to the networking protocol can also be exploited to generate fingerprint, such as TCP initial window size, IP header ID sequence generation~\cite{greenwald2007toward,smart2000defeating,greenwald2007understanding}.

\textbf{Hardware Fingerprint} In terms of hardware feature, a lot of works have been devoted to identifying the devices by exploiting minute differences of the signal produced by the component of the phone. For example, wireless NIC can be distinguished by exploiting feature from RF signal emitted by the transmitter~\cite{ureten2007wireless,franklin2006passive,desmond2008identifying,arackaparambil2010reliability,bonne2007implications,brik2008wireless}. However they cannot be promoted to Internet tracking usage, since there may be no direct physical link between user and tracer. Data collected from accelerometer can also be used to distinguish user in ~\cite{dey2014accelprint} with coarse precision without active stimulation. Photos taken by cameras can also be distinguished by pattern and noise~\cite{lukas2006digital}.

Scheme proposed in~\cite{das2014fingerprinting} also leveraged feature of speaker embedded in the phone to identify users. However, they haven't pointed out how large scale their scheme can be applied to. Besides, the robustness of the cepstral feature is not evaluated, which casts doubt on the feasibility of long-term tracking. What's more, no practical method has been postulated in his scheme in terms of hiding the identification process, while playing a clip of audible music as stimulation will inevitably attract user's attention.

\textbf{Location Stealing} Many researchers have also focused on position stealing method in android devices without corresponding permission. Zhou et al. have studied how to infer the location with public information provided by android without special permission in ~\cite{zhou2013identity}. Han et al. postulated that accelerometers in smartphones can be utilized to infer location in ~\cite{han2012accomplice}. Lester et al stated in ~\cite{lester2004you} that techniques have been found to determine if two phones are being carried by the same person. In ~\cite{nguyen2012probabilistic}, the author have raised a kind of probabilistic method for positioning to mobile devices in the pocket without GPS information.

\section{conclusion}
\label{sec:conclusion}

This paper exhibited that there are differences between speaker individuals of the smart phone, which is reflected on the differences between the response curves. It is the differences that enable applications to generate unique identity according to response curve. The identity is proved to be eligible as a kind of long-term tracking proof, because of its' stability. The identity is also proved to be entropy sufficient to incorporate all the phones in the world. In terms of anti-interference, both practical experiments and theoretical analysis are conducted to tell that the scheme works in commonly seen occasion except what with annoying high power noise. Besides the identity, more seriously, the location of the device may be exposed accompanied with, resulted from the narrow broadcasting range of the sound wave. To calculate the error rate, we analyzed the distribution model of the similarity, which is calculated by fitting the similarity between identities to some probabilistic model and choosing the most overlapped one. We decide the entropy according to size of the distinguishable device pool calculated by the error rate.

%
%%\input{modelProbability}
%
%\input{impl}
%
%\input{eval}
%
%\input{application}
%
%\input{discuss}
%
%\input{related}
%
%\input{conclusion}

%\flushend
% If you are using bibtex:

%\bibliographystyle{plain}
\bibliographystyle{abbrv}
\bibliography{main}

\end{document}